\documentclass[preprint,showpacs,preprintnumbers,amsmath,amssymb]{revtex4-1}

\usepackage{graphicx}% Include figure files
\usepackage{dcolumn}% Align table columns on decimal point
\usepackage{bm}% bold math
\begin{document}

\title{The Widom line and noise power spectral analysis of a 
supercritical fluid}

\author{Sungho~Han}
\affiliation{Department of Physics and Astronomy, University of California, 
Irvine, California 92697}

\author{Clare~C.~Yu}
\affiliation{Department of Physics and Astronomy, University of California, 
Irvine, California 92697}

\date{\today -- lvwidom.tex}

\begin{abstract}
We have performed extensive molecular dynamics simulations to study noise power spectra 
of density and potential energy fluctuations of a Lennard-Jones model of a fluid 
in the supercritical region. Emanating from the liquid-vapor critical point, there is 
a locus of isobaric specific heat maxima, called the Widom line, which
is often regarded as an extension of the liquid-vapor coexistence line.
Our simulation results show
that the noise power spectrum of the density fluctuations on the Widom 
line of the liquid-vapor transition exhibits three distinct $1/f^{\gamma}$ 
behaviors with exponents $\gamma$= 0, 1.2 and 2, depending on the frequency $f$. 
We find that the intermediate frequency region 
with an exponent $\gamma \sim$ 1 appears as the temperature approaches the Widom 
temperature from above or below. 
On the other hand, we do not find three distinct regions of $1/f^{\gamma}$ in the power 
spectrum of the potential energy fluctuations on the Widom line. Furthermore, we 
find that the power spectra of both the density and potential energy fluctuations 
at low frequency have a maximum on the Widom line, suggesting that the noise power 
can provide an alternative signature of the Widom line.
\end{abstract}


\maketitle

\section{\label{sec:intro}Introduction}
In a typical pressure-temperature $(P-T)$ phase diagram for a fluid, there is a first
order phase transition line between the liquid and the vapor phases that terminates at a
critical point~\cite{stanley}. Beyond this critical point lies the supercritical regime where
one can go continuously between the liquid and vapor phases.  In this region 
thermodynamic response functions such as the isobaric specific heat $c_p$, the 
isothermal compressibility $\kappa_{T}$ and the 
thermal expansion coefficient $\alpha$ do not monotonically increase or decrease
with $P$ and $T$. Rather they exhibit a line of maxima~\cite{nishikawa, brazjcp, brazjpcb}
emanating from the critical point that is known as the Widom
line~\cite{xu} which can be thought of as an extension of the liquid-vapor phase boundary.

More generally, a Widom line is regarded as an extension of a first order phase 
boundary that terminates
at a critical point. It is marked by maxima in the thermodynamic response functions. The
most studied Widom line is in supercooled water where a first order liquid-liquid 
transition has been proposed between a high-density liquid and a low-density 
liquid phase~\cite{poole, mishima, soper}. This phase boundary has been
hypothesized to end in a 
critical point from which the Widom line emanates~\cite{poole, mishima, xu}.
This critical point cannot be directly investigated experimentally
because it is usurped by
the spontaneous crystallization of water. Experimental searches \cite{dliu,liu}
for this critical point have involved putting water in a confined geometry
to avoid spontaneous crystallization \cite{kumarpre2005, brovchenko, han2008, han2009}.
However, so far there has been no direct experimental evidence of this critical point. 
Therefore, studies focused on the Widom line have been based on the belief 
that a Widom line is associated with the existence of a critical point
\cite{nishikawa, brazjpcb}. 
These studies have found a rich and complex behavior. For
example, along the Widom line in supercooled water, experiments have found 
a fragile-to-strong dynamic transition~\cite{liu, chen}, 
a sharp change in the proton chemical shift which is a measure of the 
local order in confined water~\cite{malla}, and a density minimum of
water \cite{dliu}. Simulations of supercooled water have found a 
breakdown of the Stokes-Einstein relation \cite{kumar2007} and a fragile-to-strong dynamic crossover~\cite{xu, pgallo} along the Widom line.
Simulations of biological macromolecules with hydration shells find that the
glass transition of the macromolecules coincides with water molecules in
the hydration shell being on the Widom line \cite{kumar2006}.

Given the interesting behavior associated with the liquid-liquid phase transition of
supercooled water,
we decided to focus on the liquid-vapor Widom line which has been largely neglected.
Experiments have found that the liquid-vapor Widom line significantly affects dynamic 
quantities as well as static quantities. X-ray diffraction measurements of
the structure factor $S(Q)$ of fluid argon
have found a liquid-like phase with high atomic correlation 
as the Widom line is approached with decreasing pressure
to something intermediate between a liquid and a gas with
low atomic correlation \cite{santoro}. (The amount of atomic correlation
is indicated by the first sharp diffraction peak in $S(Q)$ versus $Q$.)
Inelastic X-ray scattering experiments 
on the sound velocity in fluid oxygen find liquid-like 
behavior above the Widom line \cite{gorelli}, and, for the case of fluid argon, a sound
velocity that increases with wavevector at pressures in the liquid-like 
phase above the Widom line but not below it \cite{simeoni}. 
Recent theoretical simulations of a Lennard-Jones (LJ) fluid and a van der Waals fluid have 
found maxima in $c_p$, $\kappa_{T}$, $\alpha$, and density fluctuations along the Widom 
line~\cite{brazjcp, brazjpcb, han2011}. 

In this paper we use noise spectra to study the liquid-vapor Widom line.
Noise spectra have been used to probe first and second order phase transitions 
where it has been found that the low frequency noise is a maximum at the phase 
transition~\cite{zhichen}. In addition, in the vicinity of second order transitions 
the noise goes as $1/f^{\gamma}$ ($f$: frequency) where the exponent $\gamma$ can be related to 
the critical exponents~\cite{dAuiac, lauritsen, leung, zhichen}. Motivated 
by this, we decided to probe the supercritical region and the Widom line with the noise
power spectra of fluctuations in the density and the total potential energy generated from
3D molecular dynamics (MD) simulations of an LJ fluid. Along the
Widom line we find that the noise power spectrum $S_{\rho}$ of the 
density $\rho$ goes as $1/f^{\gamma}$ where the exponent
$\gamma =0$ (white noise) at low frequencies, $\gamma=1.2$ at intermediate 
frequencies, and $\gamma = 2$ at high frequencies. The behavior at intermediate
frequencies is indicative of a distribution of relaxation times \cite{dutta}. 
Away from the Widom line, there is no intermediate regime,
only white noise at low frequencies and $\gamma = 2$ at high frequencies.
At low frequencies where there is white noise in both
the density and the potential energy, the magnitude of the low frequency noise is a 
maximum along the Widom line. This is similar to the maximum in 
the low frequency noise in the energy and magnetization that was
found at the second order phase transition temperature of the 2D Ising 
ferromagnetic Ising model and at the first order transition of the 2D 5-state
Potts model \cite{zhichen}. A maximum in the low frequency noise is consistent with 
the maxima of the thermodynamic response functions at 
the phase transitions and along the Widom line.
Here we present MD simulation studies of the relation between 
the Widom line and the power spectra of the density and potential energy fluctuations 
of the supercritical LJ fluid.

This paper is organized as follows.
In Sec.~\ref{sec:methods} we describe our MD simulation of an LJ fluid in detail.
In Sec.~\ref{sec:results1} we present the results of our
calculations of thermodynamic quantities.
We show the pressure-temperature and density-temperature phase diagrams, and 
we calculate the thermodynamic response functions, such as the thermal 
expansion coefficient and the isobaric specific heat, to find the location of 
the Widom line in the phase diagram.
We also calculate the standard deviations of the density and potential energy
distributions, and find that they have their maximum value along the Widom line.
In Sec.~\ref{sec:results2} we present the noise power spectra of 
the density and potential energy fluctuations of the supercritical LJ fluid.
We find $1/f$ noise in the density fluctuations in the intermediate frequency
range around the Widom line in addition to white noise at low frequency and 
$1/f^{2}$ noise at high frequency. We also show that the power spectra of 
the density and potential energy fluctuations at low frequency 
have a maximum along the Widom line.
In Sec.~\ref{sec:summary} we summarize our results. 
       
\section{\label{sec:methods}Methods: Molecular Dynamics Simulations
of a Lennard-Jones Fluid}

The LJ interaction has
been commonly used in simulations to model fluids and in particular, 
to study the liquid-vapor phase transition
~\cite{hansen, smit, frenkel2002, wilding1995, yliu}.
The LJ interparticle potential has a combination of a 
short-ranged repulsive core and a long-ranged attractive tail: 
\begin{eqnarray}
U_{\textrm{LJ}}(r) = 4 \epsilon 
\bigg[ \bigg( \frac{\sigma}{r} \bigg)^{12} - \bigg( \frac{\sigma}{r} \bigg)^{6} \bigg].
\end{eqnarray}
Here the LJ parameters $\epsilon$ and $\sigma$ represent the energy and 
length scales, respectively, and $r$ is the distance between two particles.
Monte Carlo simulation studies~\cite{smit, frenkel2002} of LJ fluids have shown that 
the calculation of the liquid-vapor critical point is sensitive to how 
the LJ interaction is truncated at large distances.
For the full LJ interaction without truncation, the estimated 
liquid-vapor critical point is at $T^{*}_{\textrm{c}}$ $\simeq$ 1.316 
and $\rho^{*}_{\textrm{c}}$ $\simeq$ 0.304 in reduced units
(see below)~\cite{smit}.
For the truncated and shifted LJ interaction, the estimated liquid-vapor 
critical point is located at $T^{*}_{\textrm{c}}$ $\simeq$ 1.085 and 
$\rho^{*}_{\textrm{c}}$ $\simeq$ 0.317~\cite{smit, frenkel2002}.

We truncated the LJ interparticle potential at a cutoff radius 
$r_{\textrm{c}}$. In order to have the interparticle potential and force be
continuous at $r=r_{\textrm{c}}$, we truncated and then shifted the LJ 
interparticle potential~\cite{rapaport}:
\begin{eqnarray}
U(r) = \left\{\begin{array}{ll}U_{\textrm{LJ}}(r) - U_{\textrm{LJ}}(r_{\textrm{c}}) - {\displaystyle \frac{dU_{\textrm{LJ}}(r)}{dr}}\bigg|_{r=r_{\textrm{c}}} \cdot (r-r_{\textrm{c}}) & \quad 
\textrm{for $r \le r_{\textrm{c}}$}\\
0 & \quad \textrm{for $r > r_{\textrm{c}}$}.
\end{array} \right.
\end{eqnarray}
To reduce the deviation of $U(r)$ from the original LJ interparticle potential,
we used a larger cutoff radius $r_{\textrm{c}} = 4.0 \sigma$ than the 
usual cutoff radius $r_{\textrm{c}} =2.5 \sigma$~\cite{rapaport}.

Throughout this work we use reduced MD dimensionless units with the energy,
length and mass scales set by $\epsilon$, $\sigma$, and mass $m$,
respectively. Thus, the reduced length $r^{*}=r/\sigma$, energy 
$E^{*}= E/ \epsilon$, time $t^{*}=t/ \sqrt{m \sigma^{2}/\epsilon}$, 
temperature $T^{*}=k_\textrm{B} T/\epsilon$, volume $V^{*}=V/ \sigma^{3}$, 
number density $\rho^{*}=\rho\sigma^{3}$, pressure 
$P^{*}=P\sigma^{3}/ \epsilon$, specific heat  
$c^{*}_{P} = c_{P}/k_{\textrm{B}}$ and thermal expansion coefficient 
$\alpha^{*} =\alpha\epsilon/ k_{\textrm{B}}$ where $k_{\textrm{B}}$ is
Boltzmann's constant ~\cite{rapaport}. 

To study the Widom line and power spectra of supercritical LJ fluids,
we performed extensive MD simulations of 
an LJ fluid in ensembles with $NPT$ kept fixed where $N=1728$ particles,
$P$ is the pressure and $T$ is the temperature. 
The simulations were done at 33 temperatures $T^{*}$ from 1.00 up to 1.64 
with temperature increments of $\Delta T^{*} =0.02$ and for 40 pressures 
$P^{*}$ from 0.01 up to 0.4 with pressure increments of $\Delta P^{*} = 0.01$.
In the initial configuration the particles are arranged in a cubic lattice, then 
the temperature is raised to a very high temperature to melt the lattice, and then
the system is brought to the desired temperature, and $2 \times 10^{6}$ MD steps 
(or equivalently, $t^{*}$ = $10^{4}$) used to equilibrate the system, 
followed by $4 \times 10^{6}$ MD steps (or equivalently, 
$t^{*}$ = $2 \times 10^{4}$) to collect equilibrium data.
The number of steps need to achieve equilibrium was determined by seeing
how many steps were needed for the mean potential energy to become
time independent. Particle motions were updated at each time step 
$\Delta t^{*}$ = 0.005. We implemented the Berendsen thermostat and barostat 
to keep the temperature and pressure constant~\cite{berendsen}.   
We used periodic boundary conditions in the $x$, $y$ and $z$ directions.
Since the system is self-averaging, we only needed to do one run for any
given set of conditions.

\section{\label{sec:results1} Results: Thermodynamics}

\begin{figure}
\begin{center}
\includegraphics[width=0.8 \textwidth]{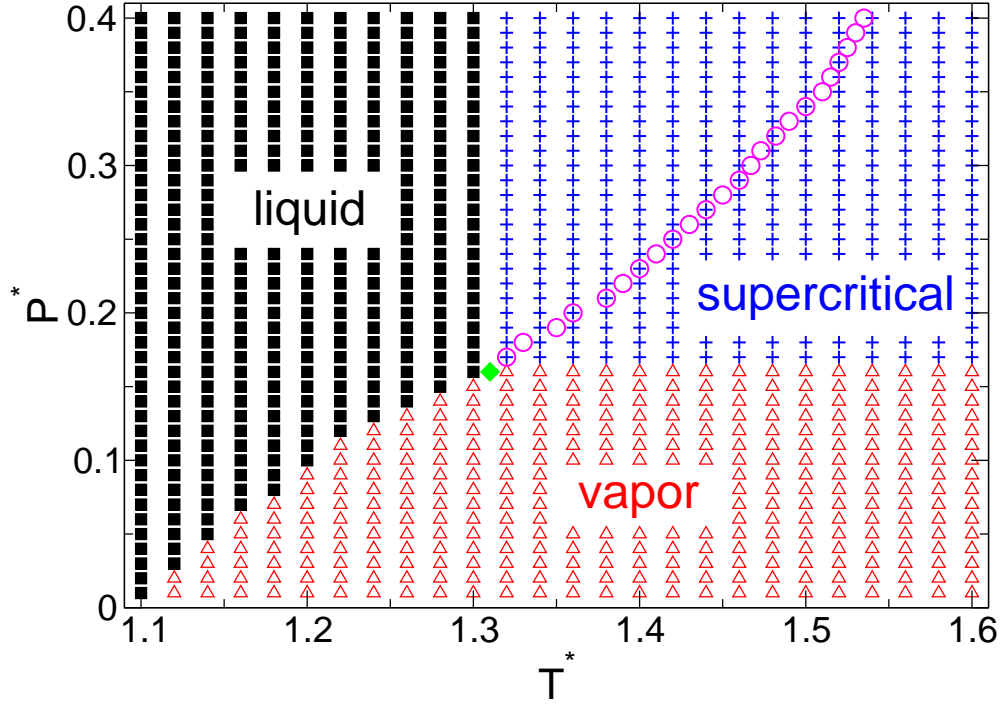}
\caption{\label{fig:ptdiagram} (Color online) Pressure-temperature ($P^{*}-T^{*}$) 
phase diagram of the liquid, vapor and supercritical fluid phases. 
The liquid and vapor phases are denoted by black squares and red triangles, 
respectively. The supercritical fluid phase is denoted by blue crosses. 
The liquid-vapor critical point estimated from our simulation results 
and Ref.~\cite{smit} is denoted by a green diamond. The approximate 
location of the liquid-vapor critical point in the $P^{*}-T^{*}$ diagram 
is $T_{\textrm{c}}^{*} \simeq 1.305$ and $P_{\textrm{c}}^{*} \simeq 0.16$. 
Circles represent the Widom line emanating from the liquid-vapor critical point. 
The Widom line is the locus of isobaric specific heat maxima $c_{\textrm{p}}$ 
estimated from Fig.~\ref{fig:supercp}. The Widom line is a 
continuous extension of the liquid-vapor phase boundary. 
All points here represent phase points we simulated.}
\end{center}
\end{figure}

\begin{figure}
\begin{center}
\includegraphics[width=0.8 \textwidth]{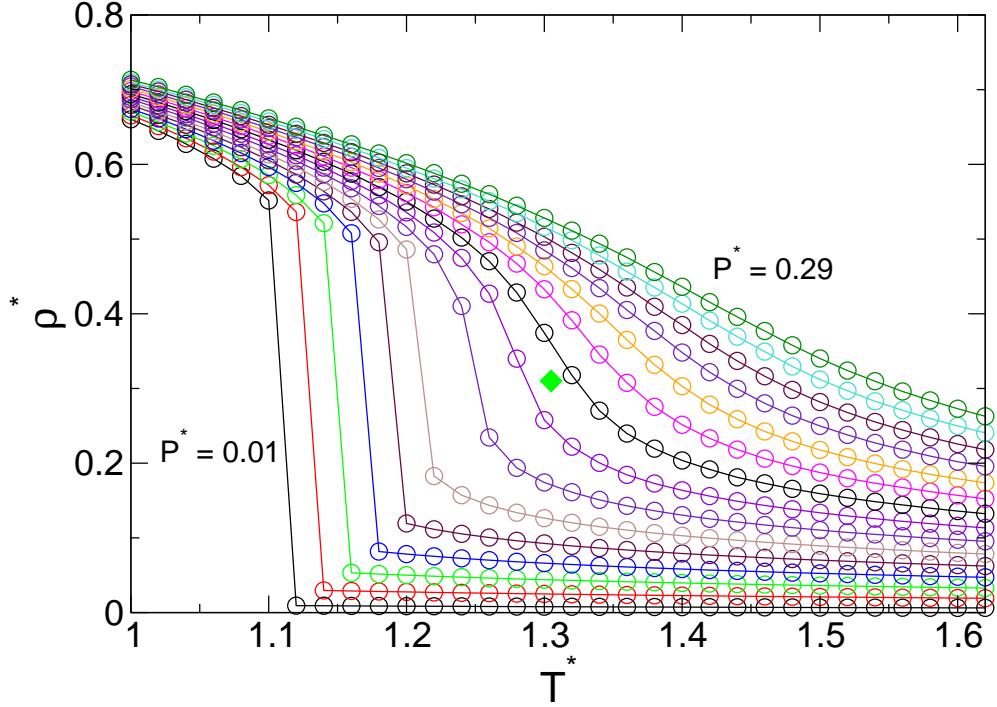}
\caption{\label{fig:dendiff} (Color online) Isobaric trajectories in the density-temperature ($\rho^{*}-T^{*}$) phase diagram from pressure $P^{*}$ = 0.01 up to $P^{*}$ = 0.29 
in increments of $\Delta P^{*}$ = 0.02. A green diamond represents the estimated liquid-vapor critical point. Below the critical point, when the system crosses the transition temperature, $\rho^{*}$ shows a discontinuous jump at the transition temperature. Above the critical point, $\rho^{*}$ shows the continuous change over the entire temperature range 
that we investigated.}
\end{center}
\end{figure}

\begin{figure}
\begin{center}
\includegraphics[width=0.8 \textwidth]{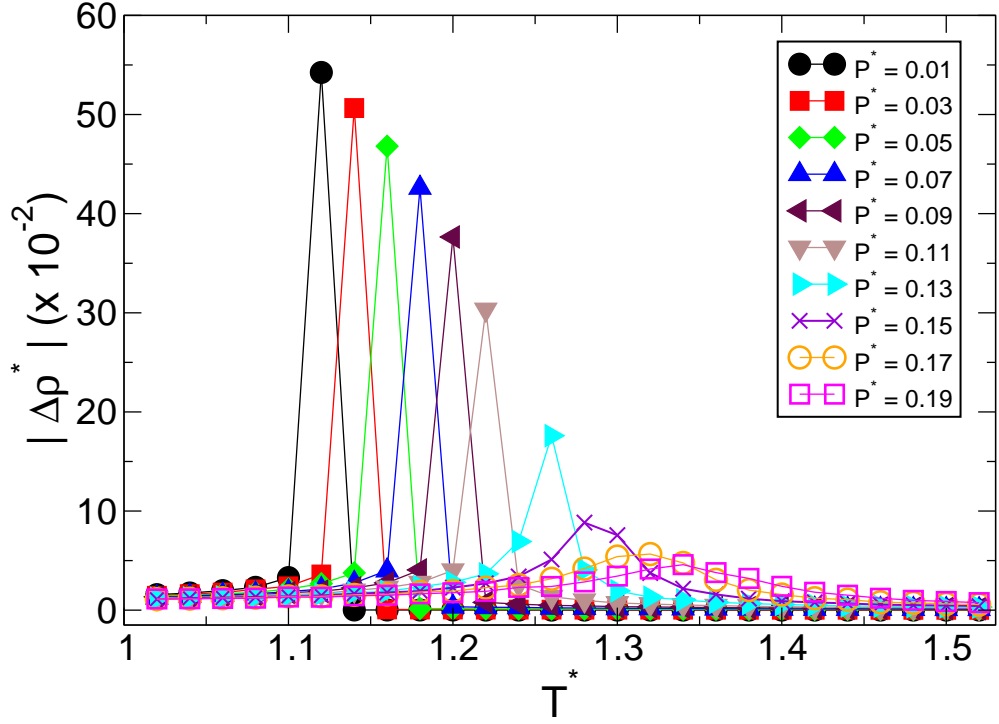}
\caption{\label{fig:deltarhobelow} (Color online) The absolute value of the 
density difference $| \Delta \rho^{*} |$ as a function of temperature $T^{*}$ 
for different pressures in the subcritical and supercritical regions. 
The absolute value of the density difference is defined by 
$| \Delta \rho^{*} (T^{*}_{i}) | \equiv |\rho^{*}(T^{*}_{i}) - \rho^{*}(T^{*}_{i-1}) |$, 
which is different from the definition of the typical order parameter of the 
liquid-vapor phase transition 
$\Delta \rho^{*} \equiv \rho_{\textrm{liquid}}^{*} - \rho_{\textrm{vapor}}^{*}$. 
The approximate location of the liquid-vapor critical point is 
$T_{\textrm{c}}^{*} \simeq$ 1.305 and $P_{\textrm{c}}^{*} \simeq$ 0.16. 
Here we used the temperature step 
$\Delta T^{*}_{i}\, (= T^{*}_{i} - T^{*}_{i-1}) = 0.02$. }
\end{center}
\end{figure}

\begin{figure}
\begin{center}
\includegraphics[width=0.8 \textwidth]{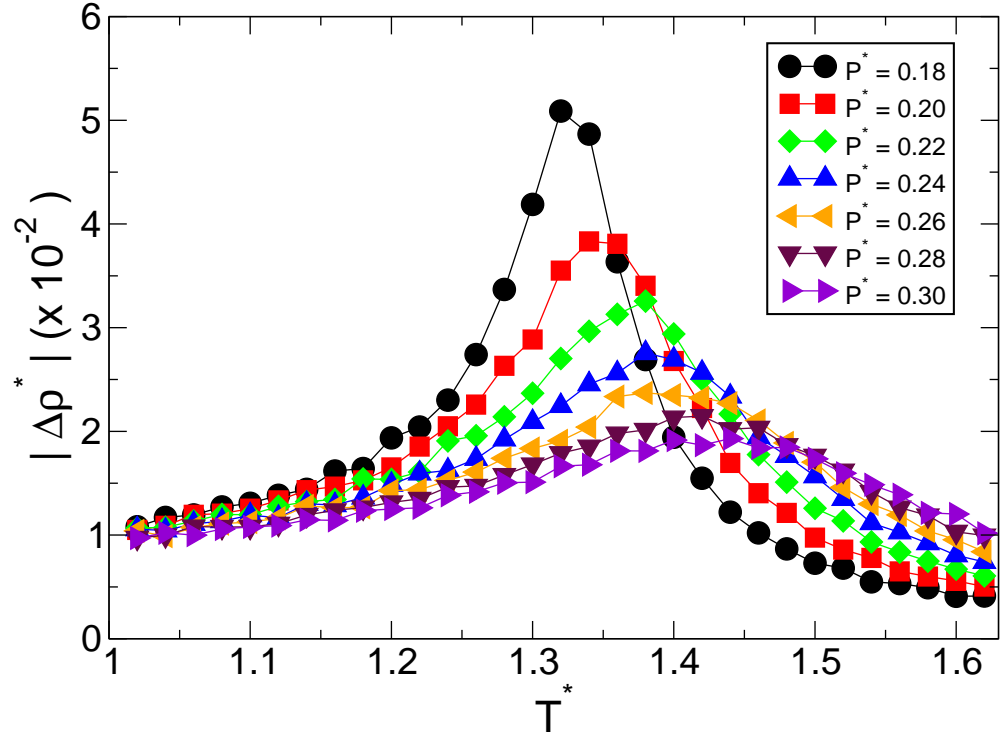}
\caption{\label{fig:superdeltarho} (Color online) Absolute value of the density 
difference $| \Delta \rho^{*} |$ versus temperature $T^{*}$ for 
different pressures in the supercritical region. The magnitude of the peaks decreases 
as the system moves away from the liquid-vapor critical point. 
At higher $P^{*}$, the location of the peak moves toward higher $T^{*}$.}
\end{center}
\end{figure}

The $P^{*}-T^{*}$ phase diagram is shown in Figure \ref{fig:ptdiagram}.
The first order phase boundary between the liquid and vapor phases terminates 
at the liquid-vapor critical point 
($T^{*}_{\textrm{c}} \simeq$ 1.305 and $P^{*}_{\textrm{c}} \simeq$ 0.16).
The Widom line continues the liquid-vapor phase boundary beyond the 
critical point. In the subcritical region where there is a first order
phase transition, the density $\rho^{*}$ along the isobaric path abruptly 
changes when $T^{*}$ crosses the transition temperature 
(Fig.~\ref{fig:dendiff}). However, $\rho^{*}$ in the supercritical region 
continuously changes as a function of $T^{*}$. These different behaviors 
of $\rho^{*}$ as a function of $T^{*}$ are clear when we calculate the 
absolute density difference between two adjacent temperatures 
$| \Delta \rho^{*} (T^{*}_{i})| \equiv 
|\rho^{*}(T^{*}_{i}) - \rho^{*}(T^{*}_{i-1})|$ 
(Figs.~\ref{fig:deltarhobelow} and \ref{fig:superdeltarho}). Note that 
this differs from the order parameter of the liquid-vapor phase 
transition $\Delta \rho \equiv \rho_{\textrm{liquid}} - \rho_{\textrm{vapor}}$.
When there is a discontinuous drop in $\rho^{*}(T^{*})$ at the transition 
temperature in the subcritical region, $|\Delta \rho^{*}|$ shows a spike 
when $T^{*}$ crosses the same transition temperature.
In the supercritical region, as shown in Fig.~\ref{fig:superdeltarho}, 
$|\Delta \rho^{*}|$ shows a maximum at a given temperature, whereas 
$\rho^{*}$ continuously changes without any abrupt jump.

The maximum in $|\Delta \rho^{*}|$ is associated with a maximum in $d\rho / dT$
that, as we will see later, is associated with the Widom
line. Similar behavior has been found in supercooled water.
As water is cooled, the density reaches a well known maximum at 4 $^{\circ}$C.
As the temperature continues to decrease into the supercooled regime,
the density decreases until it reaches a minimum as revealed by recent
inelastic neutron scattering experiments \cite{dliu}. With further
cooling the density increases in the same way that simple liquids do upon cooling
~\cite{debene}.
Between the density maximum and minimum, there is a maximum in $d\rho / dT$
which coincides with the Widom line of the liquid-liquid transition~\cite{dliu}.
Therefore, as we will see later, we expect that the temperature of the maximum 
in $|\Delta \rho^{*}|$ corresponds to the Widom temperature $T^{*}_{\textrm{w}}$.  

\begin{figure}
\begin{center}
\includegraphics[width=0.8 \textwidth]{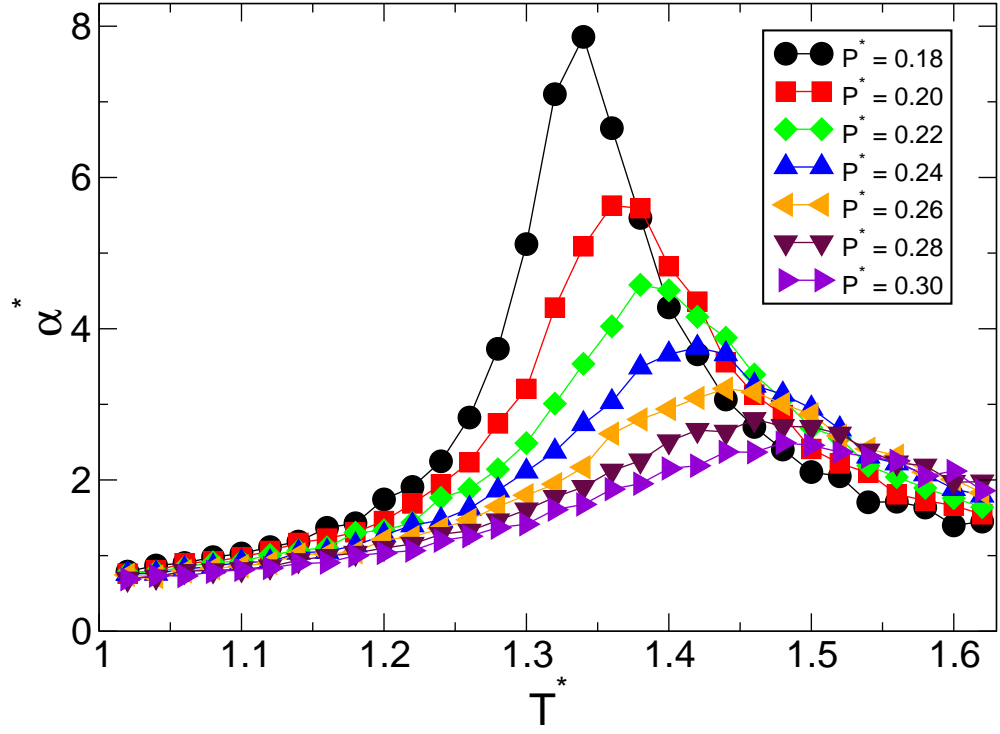}
\caption{\label{fig:superthermal} (Color online) Thermal expansion 
coefficient $\alpha^{*}$ versus temperature $T^{*}$ for different pressures 
in the supercritical region. As $P^{*}$ increases above the critical pressure $P^{*}_{\textrm{c}}$, the magnitude of the peak in $\alpha^{*}$ decreases. At higher $P^{*}$, the location of the peak moves toward higher $T^{*}$.}
\end{center}
\end{figure}

To find the Widom line, we calculate the thermal expansion coefficient 
$\alpha$ defined as~\cite{stanley}
\begin{eqnarray}
\alpha &=& \frac{1}{V} \bigg( \frac{\partial V}{ \partial T}\bigg)_{P} \nonumber\\
& =& -\bigg(\frac{\partial \ln \rho}{\partial T} \bigg)_{P}.
\end{eqnarray}
Generally, $\alpha$ is associated with fluctuations in the entropy $S$ and volume 
$V$~\cite{debene},
\begin{eqnarray}
<(\delta S \delta V)> = V k_{\textrm{B}} T \alpha.
\label{eq:alpha}
\end{eqnarray}
In Figure \ref{fig:superthermal}, we show $\alpha^{*}$ as a function of $T^{*}$ 
along the isobaric path in the supercritical region.
At low temperature, $\alpha^{*}$ initially increases gradually upon heating, and then 
starts to increase rapidly upon further heating.
As $T^{*}$ increases further, $\alpha^{*}$ finally reaches maximum, and then 
rapidly decreases.
As $P^{*}$ increases above $P^{*}_{\textrm{c}}$ (for $P^{*} > P^{*}_{\textrm{c}}$), 
the magnitude of the peak in $\alpha^{*}$ decreases, and the temperature 
of the peak in $\alpha^{*}$ moves toward higher $T^{*}$ as $P^{*}$ increases.
$\alpha^{*}$ has the same temperature behavior as $|\Delta \rho^{*}|$ at 
constant $P^{*}$. In particular, the maximum in $\alpha^{*}$ occurs at 
the same temperature as that of $|\Delta \rho^{*}|$.

\begin{figure}
\begin{center}
\includegraphics[width=0.8 \textwidth]{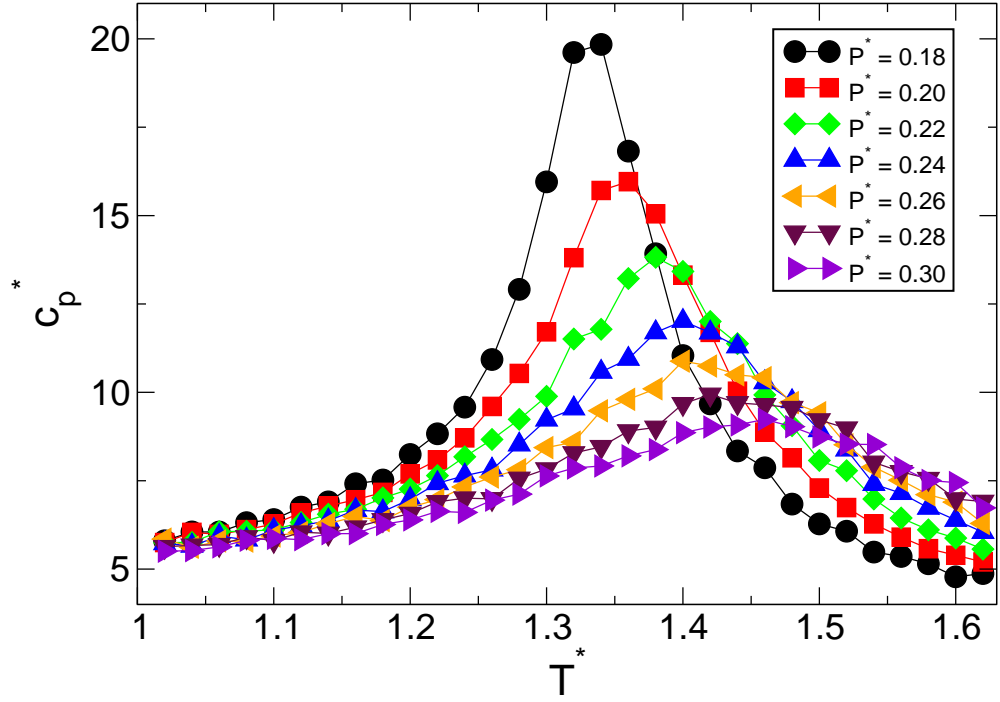}
\caption{\label{fig:supercp} (Color online) Isobaric specific heat $c^{*}_{P}$ 
versus temperature $T^{*}$ for different pressures in the supercritical region. 
As $P^{*}$ increases above $P^{*}_{\textrm{c}}$, the magnitude of the peak 
in $c^{*}_{P}$ decreases. At higher $P^{*}$, the location of the peak moves 
toward higher $T^{*}$. The behavior of $c^{*}_{P}$ is the same as that of
$|\Delta \rho^{*}|$ and $\alpha^{*}$ (see Figs.~\ref{fig:superdeltarho} 
and \ref{fig:superthermal}).}
\end{center}
\end{figure}

Next, we calculate another thermodynamic response function, the isobaric specific heat
$c_{P}$ which is defined as~\cite{stanley}
\begin{eqnarray}
c_{P} &=& \frac{1}{N} \bigg( \frac{\partial H}{\partial T} \bigg)_{P} \nonumber\\
&=& \frac{1}{N} \bigg( \frac{\partial (E + PV)}{\partial T} \bigg)_{P},
\end{eqnarray}
where $H$ is the enthalpy and $E$ is the internal energy.
Generally, $c_{P}$ is associated with fluctuations in the entropy $S$~\cite{debene},
\begin{eqnarray}
<(\delta S)^{2}> = N k_{\textrm{B}} c_{P}.
\label{eq:cp}
\end{eqnarray}
In Figure \ref{fig:supercp}, we show $c^{*}_{P}$ as a function of $T^{*}$ along 
the isobaric path in the supercritical region.
$c^{*}_{P}$ also shows the same temperature behavior that we found for 
$|\Delta \rho^{*}|$ and $\alpha^{*}$ at constant $P^{*}$ as shown in 
Figs.~\ref{fig:superdeltarho} and \ref{fig:superthermal}.
As $P^{*}$ increases from $P^{*}_{\textrm{c}}$ (for $P^{*} > P^{*}_{\textrm{c}}$), 
the magnitude of the peak in $c^{*}_{P}$ decreases, and the temperature of the 
peak in $c^{*}_{P}$ moves toward higher $T^{*}$ as $P^{*}$ increases.
The temperature of the $c^{*}_{P}$ maximum in Fig.~\ref{fig:supercp} is 
the Widom temperature $T^{*}_{\textrm{w}}$ at a given pressure
\cite{xu, kumar2006, kumar2007, kumar2008, stokely}.  
Note that of all the various response functions, the isobaric specific heat 
is usually used to define the location of the Widom line~\cite{xu, kumar2006}.
Based on our calculation of $c^{*}_{P}$, we plot the estimated locations 
of $c^{*}_{P}$ maxima (the Widom line) in Fig.~\ref{fig:ptdiagram}, 
denoted by open circles.

\begin{figure}
\begin{center}
\includegraphics[width=0.8 \textwidth]{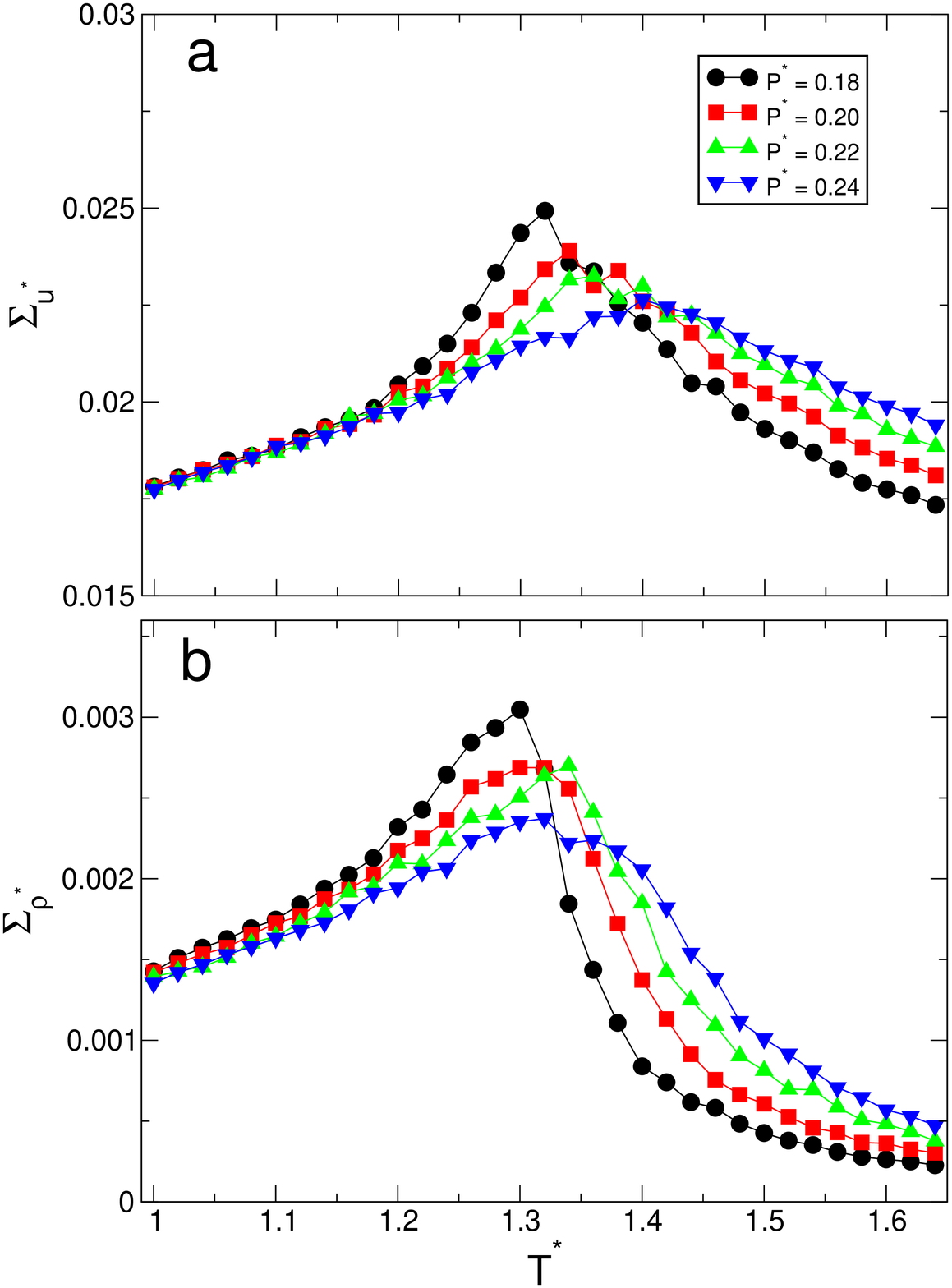}
\caption{\label{fig:var} (Color online) Standard deviations 
(a) $\Sigma_{u^{*}}$ of the potential energy $u^{*}$ per particle and 
(b) $\Sigma_{\rho^{*}}$ of density $\rho^{*}$ as a function of 
temperature $T^{*}$. Both $\Sigma_{u^{*}}$ and $\Sigma_{\rho^{*}}$ 
show maxima around the Widom line.}
\end{center}
\end{figure}

Next we investigate the standard deviation of an observable $X$ defined as
\begin{eqnarray}
\Sigma_{X} \equiv \sqrt{ \big< (X - \bar{X})^2 \big>},
\end{eqnarray}
where $\bar{X}$ is the average of $X$.
We calculate the standard deviations $\Sigma_{u^{*}}$ and $\Sigma_{\rho^{*}}$ 
of the potential energy per particle $u^{*}$ and density $\rho^{*}$, respectively.
In Figure \ref{fig:var}, we present $\Sigma_{u^{*}}$ and $\Sigma_{\rho^{*}}$ as 
a function of $T^{*}$.
The fluctuations in $u^{*}$ and $\rho^{*}$ rapidly increase as $T^{*}$ approaches 
$T^{*}_{\textrm{w}}$ from below at fixed $P^{*}$, and they reach their
maximum around $T^{*} \simeq T^{*}_{\textrm{w}}$.
After crossing $T^{*}_{\textrm{w}}$, $\Sigma_{u^{*}}$ and $\Sigma_{\rho^{*}}$ 
rapidly decrease as shown in Fig.~\ref{fig:var}.
The location in the phase diagram of the maxima in the variances agrees 
well with the location of the maxima of the thermodynamic response functions along
the Widom line. This is consistent with Eqs.~(\ref{eq:alpha}) and~(\ref{eq:cp}) 
which relate fluctuations to thermodynamic response functions.

\section{\label{sec:results2} Noise Power Spectral Analysis}
Let us set up our notation and define what we mean by noise.
Let $m(t)$ be a quantity that fluctuates in time.
Let $\delta m(t)$ be the deviation from its average value 
of some quantity $m$ at time $t$.
If the processes producing the fluctuations are stationary in time, i.e.,
translationally invariant in time, then
the autocorrelation function of the fluctuations
$\langle\delta m(t_2)\delta m(t_1)\rangle$ will be a function 
$\psi(t_2-t_1)$
of the time difference. In this case the Wiener--Khintchine theorem
can be used to relate the noise spectral density $S_m(\omega)$ to the
Fourier transform $\psi(\omega)$ of the autocorrelation 
function \cite{Kogan96}: $S_{m}(\omega)=2\psi_{m}(\omega)$
where $\omega$ is the angular frequency. In practice $S_m(\omega)$
typically is calculated by multiplying the time series $\delta m(t)$
by a windowing or envelope function so that the time series goes
smoothly to zero, Fourier transforming the result, taking the modulus
squared, and multiplying by two to obtain the noise
power \cite{Press92}. (We find that our results are not sensitive to
the choice of windowing function, so we do not use a windowing function; this  
is equivalent to a rectangular window.)

$1/f$ noise, where $\displaystyle f \, \big(= \frac{\omega}{2 \pi} \big)$ is frequency, corresponds to 
$S_m(\omega)\sim 1/\omega$. It dominates at low frequencies
and has been observed in a wide variety of systems, such as 
granular systems, molecular liquids, ionic liquids, a lattice gas model, 
and resistors~\cite{dutta, bak, weissman, jensen, sasai, milotti, 
reichhardt2003, mudi, sharma, reichhardt2007, zhichen, jeong}. 
For example, long time fluctuations of the potential energy in water and silica 
exhibiting $1/f$ spectra have been reported in computer simulation studies
\cite{sasai, mudi, sharma}. The $1/f$ power spectrum of the potential energy 
fluctuations for water is related to the non-exponential relaxation of slow 
hydrogen bond dynamics~\cite{han2009, mudi}. A study of solvation dynamics in an 
ionic liquid at room temperature has also shown $1/f$ spectral behavior~\cite{jeong}.
Fluctuations in the number of defects in a disordered two-dimensional liquid 
also exhibit a $1/f$ power spectrum at low temperatures, suggesting that 
the dynamics 
of the system is heterogeneous~\cite{reichhardt2003, reichhardt2007}. In addition 
to the relation between $1/f$ spectral behavior and dynamics, the power 
spectra can be used as a probe of phase transitions.
It has been shown that at a phase transition in classical spin systems 
(such as the Ising model and Potts model), the low frequency noise of the energy and 
magnetization fluctuations has a maximum at the transition temperature~\cite{zhichen}.

A simple way to obtain 1/f noise was given by Dutta and Horn \cite{dutta}.
We can use the relaxation time approximation to write the equation of motion for
$\delta p$:
\begin{equation}
\frac{d\delta p}{dt}=-\frac{\delta p}{\tau}
\end{equation}
where $\tau^{-1}$ is the relaxation rate. The autocorrelation function
$\psi_{p}(t)$ is given by
\begin{equation}
\psi_{p}(t)=\langle\delta p(t)\delta p(t=0)\rangle.
\end{equation}
The Fourier transform is a Lorentzian if there is just one value of $\tau$.
\begin{equation}
S_{p}(\omega)=\frac{A\tau}{1+\omega^2\tau^{2}}
\label{eqn:single}
\end{equation}
where $A$ is an overall scale factor.
If there is a broad distribution $g(\tau)$ of relaxation times, then
the Fourier transform is a sum of Lorentzians:
\begin{equation}
S_{p}(\omega)=A\int^{\tau_2}_{\tau_1}\frac{\tau}
{1+\omega^2\tau^{2}}g(\tau)d\tau
\label{eq:lorentzians}
\end{equation}
where $\tau_2$ and $\tau_1$ are the upper and lower limits 
of the distribution $g(\tau)$ and, hence, of the integral over $\tau$. If we assume that 
$g(\tau) \sim \tau^{-1}$~\cite{dutta}, then we obtain 
\begin{eqnarray}
S_{p} (\omega) \propto \frac{A}{\omega} \bigg[ \arctan \bigg( \frac{1}{\omega \tau_{2}} \bigg) - \arctan\bigg( \frac{1}{\omega \tau_{1}}\bigg)\bigg].
\label{eqn:multiple}
\end{eqnarray}
In the low frequency limit ($f \ll \tau_{2}^{-1} \ll \tau_{1}^{-1}$), 
the power spectrum 
shows $1/f^{0}$ behavior, and in the high frequency limit 
($f \gg \tau_{1}^{-1} \gg \tau_{2}^{-1}$), it shows $1/f^{2}$ behavior.
In the intermediate frequency region ($\tau_{2}^{-1} \ll f \ll \tau_{1}^{-1}$), 
the power spectrum exhibits $1/f$ behavior~\cite{dutta}.

Here we use the noise power spectra of $\rho^{*}$ and $u^{*}$ 
as a probe of the Widom line. 
The power spectrum of an observable $X(t)$ is defined as
\begin{eqnarray}
S_{X}(f^{*}) \equiv \bigg| \int X(t^{*}) e^{2\pi i f^{*} t^{*} }\,dt^{*} \bigg|^{2}.
\end{eqnarray}
There is no subtraction of the mean $\overline{X}$, so there will be a delta
function at $f=0$. $S_{X}(f)$ is otherwise the same as the noise spectrum of
the fluctuations $\delta X$. 

\subsection{Noise Spectra of Density Fluctuations}
\begin{figure}
\begin{center}
\includegraphics[width=0.8 \textwidth]{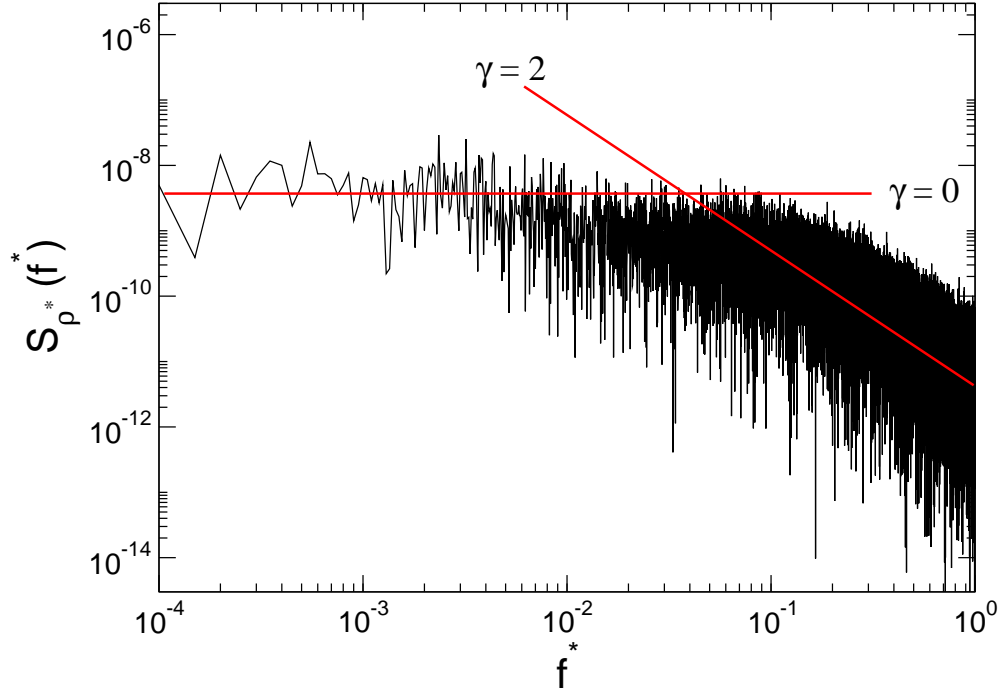}
\caption{\label{fig:rhoS1} (Color online) Noise power spectrum $S_{\rho^{*}}$ of 
the density $\rho^{*}$ as a function of frequency $f^{*}$ at 
$T^{*}$ = 1.02 ($\ll$ $T^{*}_{\textrm{w}}$) and $P^{*}$ = 0.20 in a log-log plot. 
The two solid lines are fits to $1/f^{* \gamma}$ 
with the exponents $\gamma$= 0 and 2, respectively.}
\end{center}
\end{figure}

\begin{figure}
\begin{center}
\includegraphics[width=0.8 \textwidth]{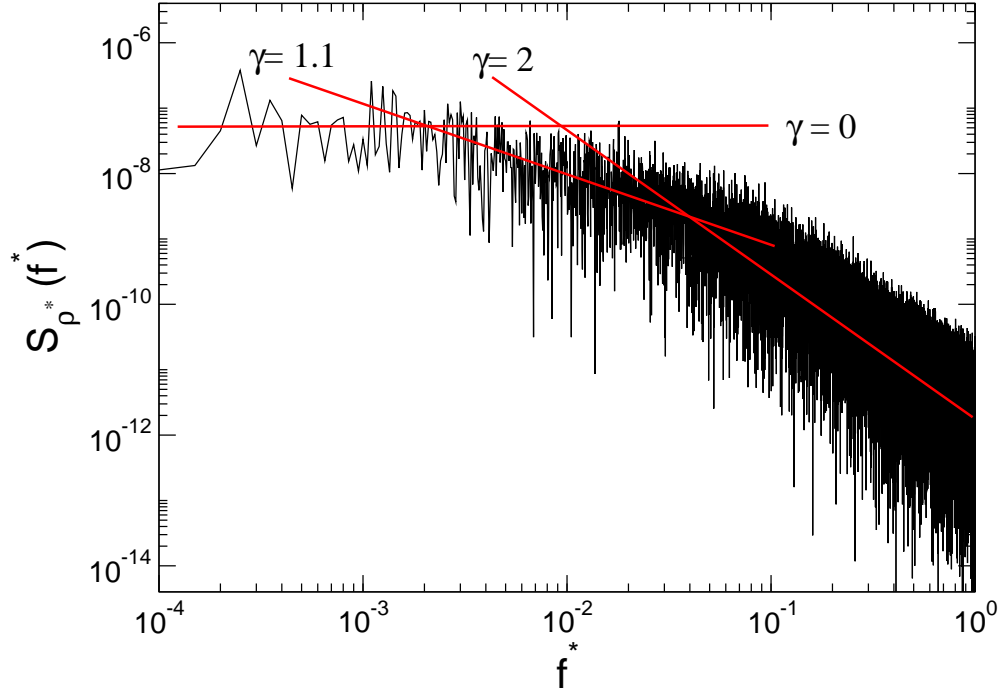}
\caption{\label{fig:rhoS4} (Color online) 
The noise power spectrum 
$S_{\rho^{*}}$ of the density $\rho^{*}$ versus
frequency $f^{*}$ at $T^{*}$ = 1.26 ($<$ $T^{*}_{\textrm{w}}$) and 
$P^{*}$ = 0.20 on a log-log plot. 
The point ($T^{*}$ = 1.26, $P^{*}$ = 0.20) is located in 
the liquid region of the phase diagram where
the response functions ($\alpha$ and $c_{P}$) change rapidly below 
$T_{\textrm{w}}$. That is, $\alpha$ and $c_{P}$ at $T^{*}$ = 1.26 rapidly 
increase as $T^{*}$ increases, as shown in Figs.~\ref{fig:superthermal} and 
\ref{fig:supercp}. The three solid lines are fits to $1/f^{* \gamma}$ with 
the exponents $\gamma$= 0, 1.1, and 2.}
\end{center}
\end{figure}

\begin{figure}
\begin{center}
\includegraphics[width=0.8 \textwidth]{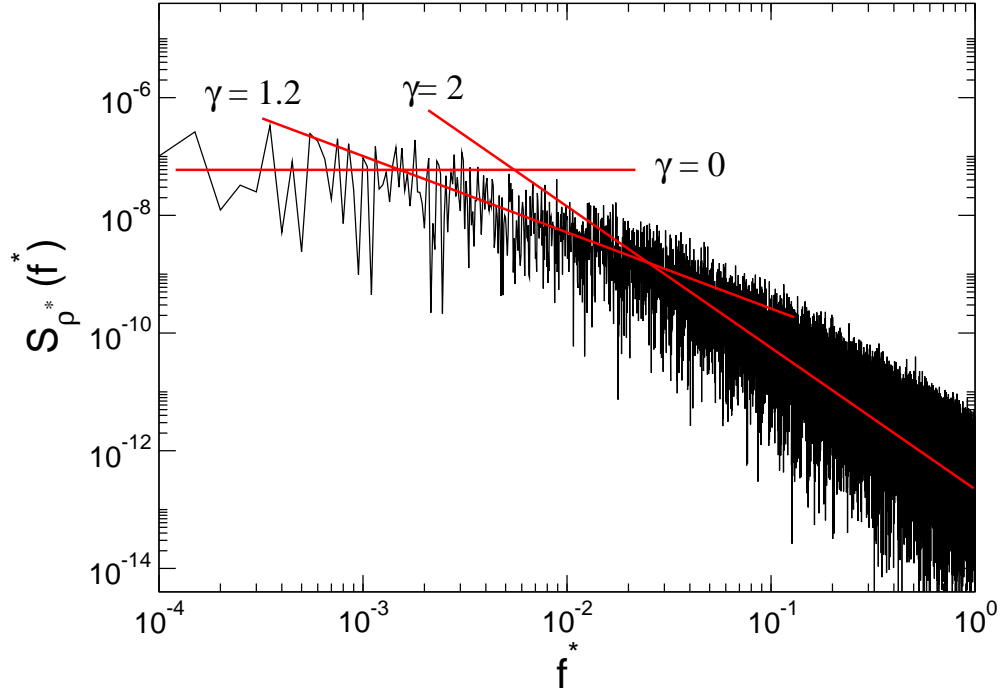}
\caption{\label{fig:rhoS2} (Color online) 
Power spectrum $S_{\rho^{*}}$ of the density $\rho^{*}$ as a function of 
frequency $f^{*}$ at $T^{*}$ = 1.36 ($\approx$ $T^{*}_{\textrm{w}}$) and 
$P^{*}$ = 0.20 on a log-log scale. The three solid lines indicate 
fits to $1/f^{* \gamma}$ with the exponents $\gamma$= 0, 1.2, and 2, respectively.}
\end{center}
\end{figure}

\begin{figure}
\begin{center}
\includegraphics[width=0.55 \textwidth]{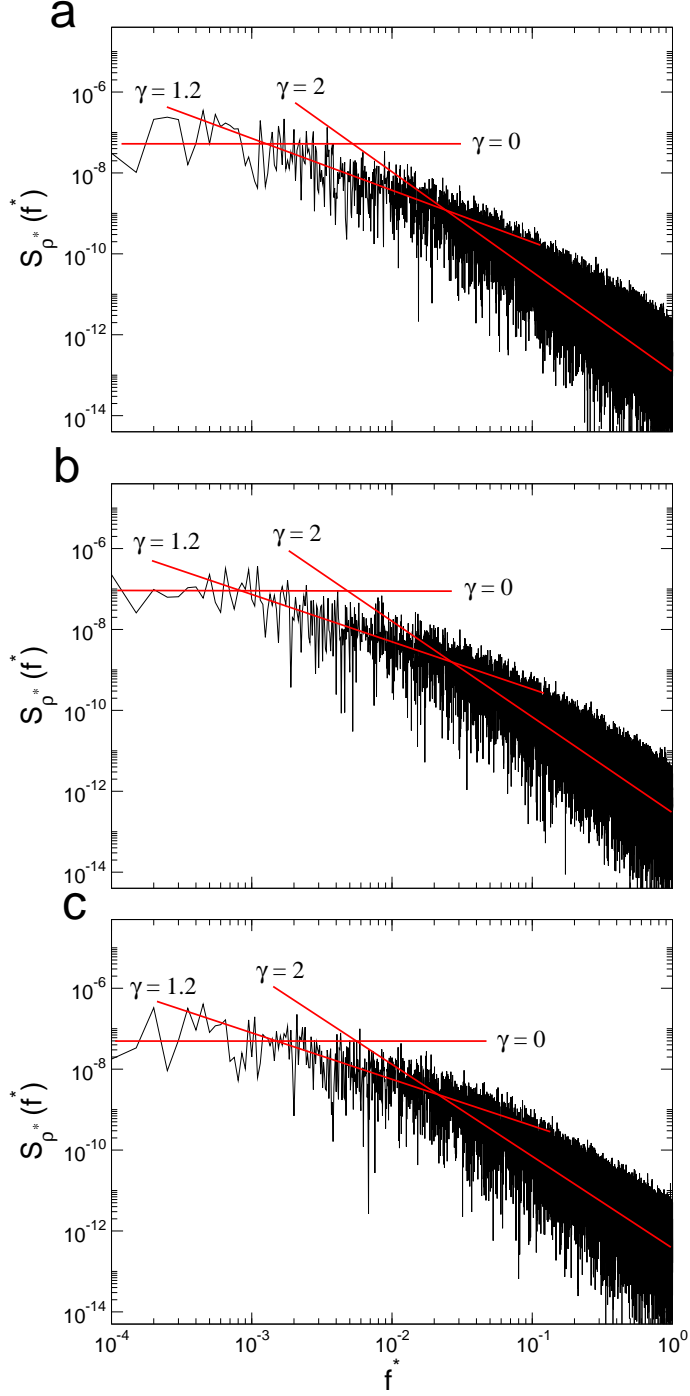}
\caption{\label{fig:rhoS6} (Color online) 
Shown on a log-log scale are power spectra $S_{\rho^{*}}$ of the 
density $\rho^{*}$ as a function of frequency $f^{*}$ at 
(a) $T^{*}$= 1.34 and $P^{*}$= 0.18, (b) $T^{*}$= 1.38 and $P^{*}$= 0.22, 
and (c) $T^{*}$= 1.40 and $P^{*}$= 0.24. These temperatures are the 
Widom temperature $T^{*}_{\textrm{w}} (P^{*})$ at a given pressure 
$P^{*}$. $S_{\rho^{*}}$ along the Widom line exhibits three distinct 
frequency regimes where $1/f^{\gamma}$ has exponents 
$\gamma$= 0, 1.2 and 2.}
\end{center}
\end{figure}

\begin{figure}
\begin{center}
\includegraphics[width=0.8 \textwidth]{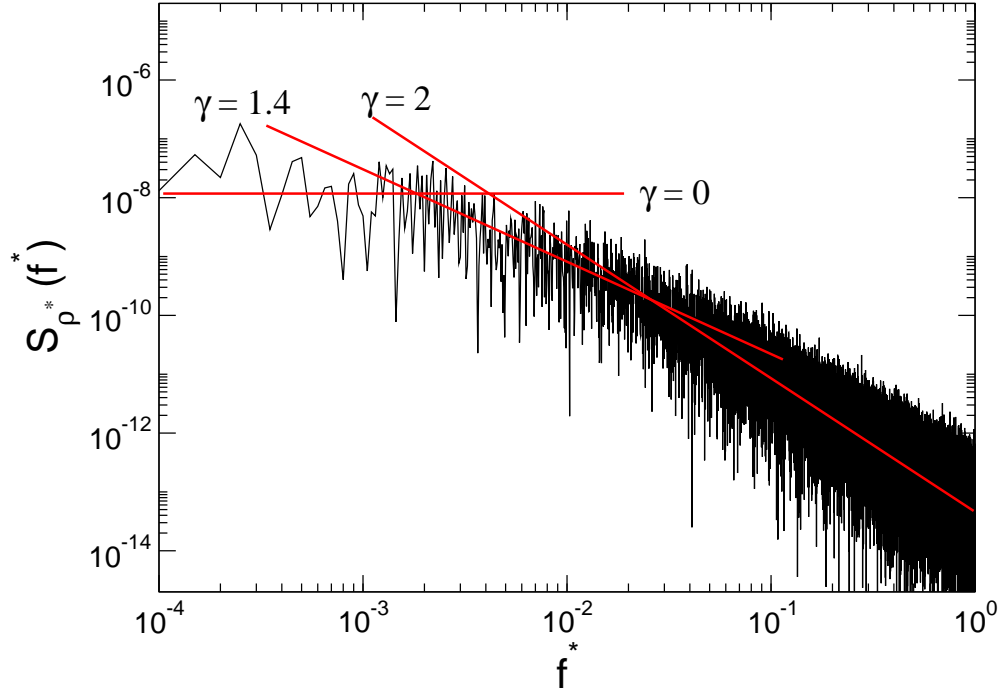}
\caption{\label{fig:rhoS5} (Color online) 
The power spectrum $S_{\rho^{*}}$ of the density $\rho^{*}$ as a function 
of frequency $f^{*}$ at $T^{*}$ = 1.44 ($>$ $T_{\textrm{w}}$) and $P^{*}$ = 0.20 
on a log-log scale. $T^{*}$ = 1.44 is located in the region 
of the phase diagram where the response functions 
($\alpha$ and $c_{P}$) change rapidly above $T_{\textrm{w}}$. 
That is, $\alpha$ and $c_{P}$ at $T^{*}$ = 1.44 rapidly decrease as $T^{*}$ 
increases, as shown in Figs.~\ref{fig:superthermal} and \ref{fig:supercp}. 
The three solid lines are fits to $1/f^{* \gamma}$ with the
exponents $\gamma$ = 0, 1.4 and 2, respectively.}
\end{center}
\end{figure}

\begin{figure}
\begin{center}
\includegraphics[width=0.8 \textwidth]{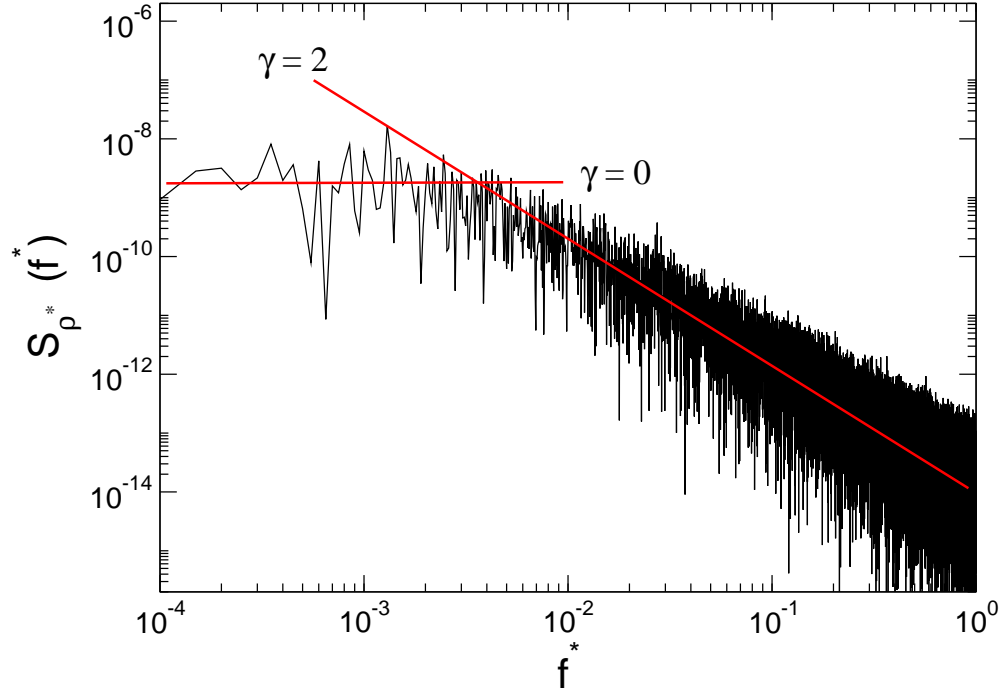}
\caption{\label{fig:rhoS3} (Color online) 
Shown on a log-log plot is the power spectrum $S_{\rho^{*}}$ of the
density $\rho^{*}$ as a function of frequency $f^{*}$ at 
$T^{*}$ = 1.58 $(\gg T^{*}_{\textrm{w}})$ and $P^{*} = 0.20$. The
two solid lines are fits to $1/f^{* \gamma}$ with the exponents 
$\gamma$= 0 and 2, respectively.}
\end{center}
\end{figure}

In Figures \ref{fig:rhoS1} to \ref{fig:rhoS3}, 
we present the power spectra $S_{\rho^{*}}$ of the density fluctuations as a 
function of frequency $f^{*}$ for different temperatures along the isobaric 
path of $P^{*}$= 0.20.
When the temperature $T^{*}$ (= 1.02) is far below the Widom temperature 
$T^{*}_{\textrm{w}}\simeq 1.36$ such that $T^{*} \ll T^{*}_{\textrm{w}}$, 
$S_{\rho^{*}}$ shows two different frequency dependences: almost flat 
($\sim 1/f^{* 0}$) at low frequencies and rapidly decreasing 
($\sim 1/f^{* 2}$) at high frequencies (see Fig.~\ref{fig:rhoS1}).
At $T^{*}=1.26$, 
the thermodynamic response functions start to change 
rapidly, and $S_{\rho^{*}}(f^{*})$ clearly exhibits three different 
frequency dependences: flat ($\sim 1/f^{* 0}$) at low frequencies, 
relatively slow decreasing ($\sim 1/f^{* 1.1}$) with increasing
frequency at intermediate frequencies, 
and rapidly decreasing ($\sim 1/f^{* 2}$) at high frequencies 
(see Fig.~\ref{fig:rhoS4}).
On the Widom line ($T^{*}= 1.36 \simeq T^{*}_{\textrm{w}}$), $S_{\rho^{*}}$, 
as shown in Fig.~\ref{fig:rhoS2}, shows the same three different frequency 
dependences as in Fig.~\ref{fig:rhoS4}: flat ($\sim 1/f^{* 0}$) at low frequencies, 
relatively slow decreasing ($\sim 1/f^{* 1.2}$) at intermediate frequencies 
and rapidly decreasing ($\sim 1/f^{* 2}$) at high frequencies.
For different $P^{*}$, $S_{\rho^{*}}(f)\sim 1/f^{* \gamma}$ at the 
corresponding Widom temperature $T^{*}_{\textrm{w}} (P^{*})$ with 
$\gamma$= 0, 1.2 and 2, respectively (see Fig.~\ref{fig:rhoS6}).
Interestingly, in the intermediate frequency region, 
$\gamma \simeq 1.2$ all along the Widom line, 
independent of $P^{*}$.

The behavior of $S_{\rho^{*}}$ at temperatures in the vicinity of
$T^{*}_{\textrm{w}}$ is similar to that seen along the Widom line.
Below but close to $T^{*}_{\textrm{w}}$ where
the thermodynamic response functions still change rapidly (as in the case 
of $T^{*}$ = 1.44 shown in Fig.~\ref{fig:rhoS5}), $S_{\rho^{*}}$ shows 
three different frequency dependences: white ($\sim 1/f^{* 0}$) at low 
frequencies, relatively slow decreasing ($\sim 1/f^{* 1.4}$) at intermediate 
frequencies and rapidly decreasing ($\sim 1/f^{* 2}$) at high frequencies.
When $T^{*}$ is far away from $T^{*}_{\textrm{w}}$, there are two frequency 
dependences: flat ($\sim 1/f^{* 0}$) at low frequencies and rapidly decreasing 
($\sim 1/f^{* 2}$) at high frequencies (see Fig.~\ref{fig:rhoS3} where 
$T^{*} = 1.58 \gg T^{*}_{\textrm{w}}$)).

When $T^{*}$ is far away from $T^{*}_{\textrm{w}}$, the density fluctuations
relax exponentially with a single relaxation time.  
As explained in the simple model of $1/f$ noise in Eqs.~(\ref{eqn:single}) 
and (\ref{eqn:multiple}), $1/f$ spectral behavior can be connected with 
a broad distribution of relaxation times. Therefore, our results imply 
that as $T^{*}$ approaches $T^{*}_{\textrm{w}}$, the 
density fluctuations are becoming more complicated and
are relaxing with a broad distribution of relaxation times.
The appearance of the $1/f$ spectral behavior near the Widom line in
a narrow frequency range might be from the fact that the Widom line of 
the liquid-vapor transition is located at very high temperatures, at which 
heterogeneous dynamics generally does not occur
\cite{reichhardt2003, reichhardt2007}, so that the distribution of 
relaxation times might have a narrow range of the values as implied by
Eq.~(\ref{eqn:multiple}). 
Note that $S_{\rho^{*}}$ along the Widom line does not provide a clear 
distinction between liquid-like and vapor-like behaviors as the recent 
x-ray scattering experiments on the velocity of sound in argon did \cite{simeoni}.

\subsection{Noise Spectra of Potential Energy Fluctuations}
\begin{figure}
\begin{center}
\includegraphics[width=0.8 \textwidth]{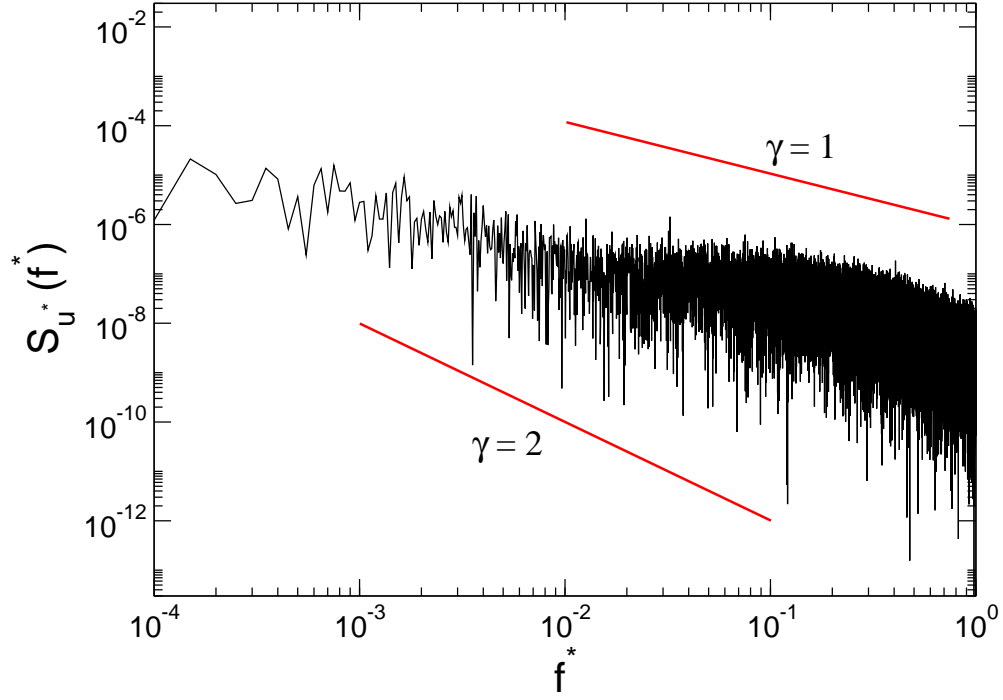}
\caption{\label{fig:peS2} (Color online) 
The power spectrum $S_{u^{*}}$ of the potential energy per particle 
$u^{*}$ as a function of frequency $f^{*}$ at $T^{*}$ = 1.34 
$(\simeq T^{*}_{\textrm{w}})$ and $P^{*}$ = 0.20 on a log-log scale. 
The two solid lines are guides to the eye and indicate $1/f^{* \gamma}$ 
with the exponents $\gamma$= 1 and 2, respectively.}
\end{center}
\end{figure}

We now examine the fluctuations in the potential energy $u^{*}$ per particle.
In contrast to $S_{\rho^{*}}$, the noise spectra $S_{u^{*}}$ of potential 
energy fluctuations does not clearly show three different frequency 
dependences on the Widom line, as shown in Fig.~\ref{fig:peS2}. $S_{u^{*}}(f)$ 
is white at low frequencies and then slowly decreases with increasing frequency.
Previous simulations of argon at $T$ = 95 K found that 
the noise spectrum of the total potential energy fluctuations is white over a wide 
range of frequencies, followed by a rapid decrease at high 
frequencies~\cite{sasai}.
Interestingly, as opposed to the total potential energy, the potential
energy fluctuations of an individual argon atom exhibit $1/f$ noise in an
intermediate frequency range ($1 \lesssim f \lesssim 10$ cm$^{-1}$)~\cite{sasai}. 
The potential energy of an argon cluster also exhibits $1/f$ noise in an 
intermediate frequency range ($f \lesssim 1$ cm$^{-1}$), indicating that 
locally the system can have a distribution of
relaxation times, e.g., at the core and surface of the argon cluster~\cite{sasai}.
The origin of the difference between the total potential energy fluctuations
of simple liquids and the
potential energy fluctuations individual atoms and clusters remains elusive
\cite{sasai}. An investigation of this difference would be interesting,
but is beyond the scope of the present work.
We should also mention that a previous computational study found that the 
long-time total potential energy 
fluctuations of water does exhibit $1/f$ noise, unlike simple liquids like argon 
\cite{sasai}. Water has a random hydrogen-bond network that has non-exponential 
relaxation processes, whereas simple liquids do not have such structures
\cite{han2009, matsumoto}.

\subsection{Block Average of Variances in Density and Potential Energy}
We are able to estimate the correlation 
time of an observable $X(t)$ by calculating the block average
\cite{frenkel2002, zhichen, rapaport, flyvbjerg, ferrenberg, yu2004}.
To calculate the block average, we take a time series of $X(t)$, divide the time
series into blocks or bins of equal size, calculate the thermodynamic
quantity such as the variance $\Sigma_{X}$ of $X$ 
for each bin, and then average over all the bins to obtain the block average. 
If the quantity is not a linear function of $X$, then the block average 
can depend on the bin size. So one can use the same time series over again
to calculate the block average with a different bin size, and see if the block
average changes with bin size. Typically, the block average will increase with
bin size before saturating at a constant value corresponding to the equilibrium
value of the quantity \cite{zhichen,yu2004}. The saturation occurs when the bin size
exceeds the correlation time~\cite{rapaport, flyvbjerg, ferrenberg}.

\begin{figure}
\begin{center}
\includegraphics[width=0.8 \textwidth]{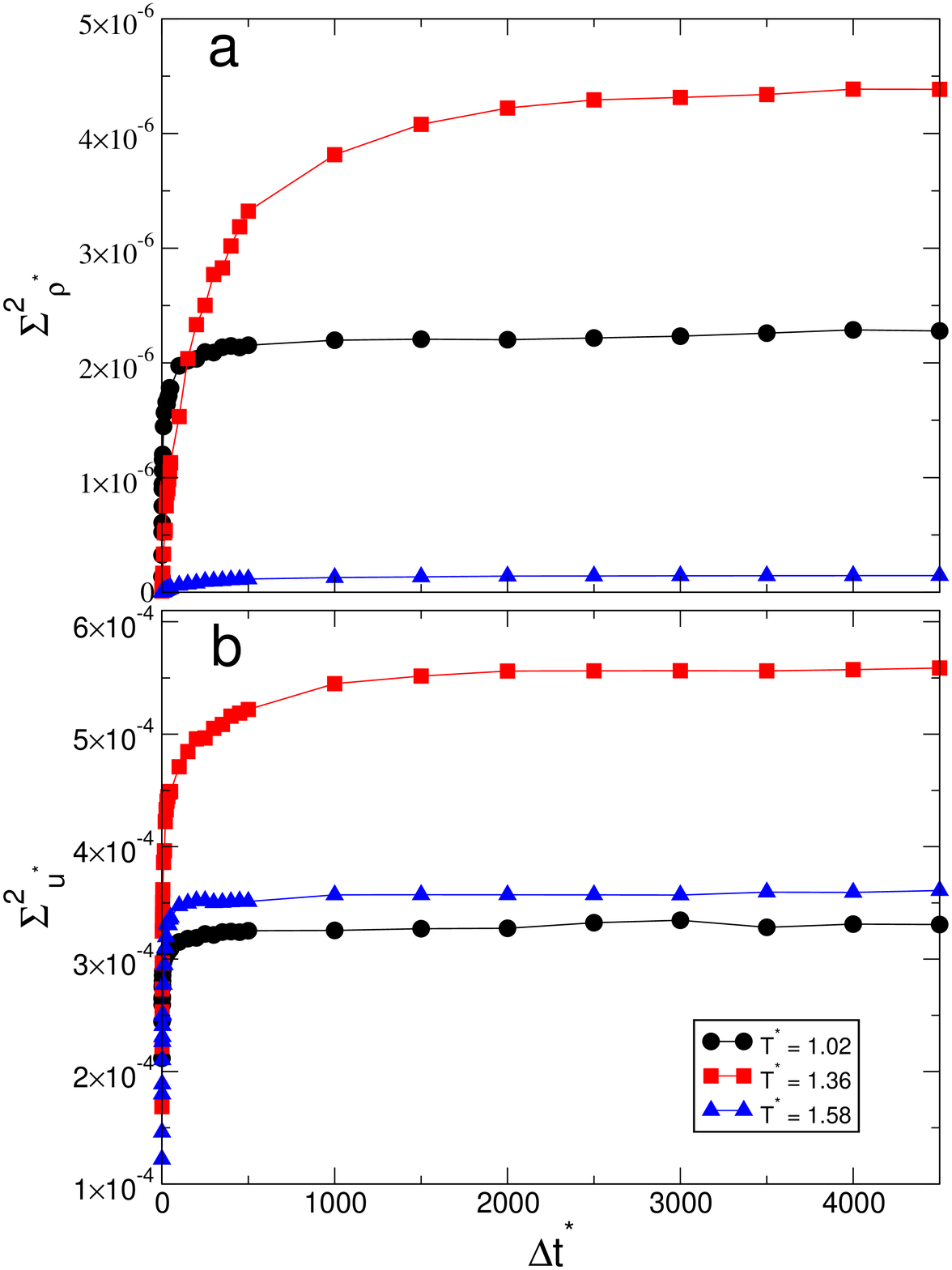}
\caption{\label{fig:block} (Color online) 
Block averages of (a) the density variance $\Sigma^{2}_{\rho^{*}}$, and 
(b) the potential energy variance $\Sigma^{2}_{u^{2}}$ as a function of the 
block size $\Delta t^{*}$ at $P^{*}$ = 0.20 for different temperatures 
$T^{*}$ = 1.02 ($< T^{*}_{\textrm{w}}$), 1.36 ($\simeq T^{*}_{\textrm{w}}$), 
and 1.58 ($> T^{*}_{\textrm{w}}$).}
\end{center}
\end{figure}

In Figure \ref{fig:block}, we calculate the block average of the variance in 
the density ($\Sigma^{2}_{\rho^{*}}$) and in the potential energy per particle 
($\Sigma^{2}_{u^{*}}$) as a function of the block size 
($\Delta t^{*}$) for different temperatures.
By comparing the block averages for three different temperatures, we find 
that the correlation times of both $\rho^{*}$ and $u^{*}$ are largest at the 
Widom temperature ($T^{*}_{\textrm{w}} \simeq 1.36$), where
$|\Delta\rho^{*}|$ and the specific heat have their maxima.
The block averages for both quantities saturate at a constant value at 
approximately $\Delta t^{*} \sim 10^{3}$. This block size indicates 
the largest correlation time and corresponds to the crossover frequency 
between the $1/f^{*0}$ and $1/f^{*}$ spectral behaviors that occurs at
approximately $f^{*} \sim 10^{-3}$.
Recent Monte Carlo studies of the classical spin systems also found that
the crossover between white noise at low frequencies and $1/f$ noise at 
higher frequencies occurred at a frequency corresponding to the inverse
of the largest correlation time \cite{zhichen}.
It is interesting to note that at the Widom temperature,
the correlation time of the density fluctuations is larger than that of 
the potential energy fluctuations.

\subsection{Maximum of Low Frequency Noise as a Signature of the Widom Line}

A maximum in the low frequency white noise can be a signature of a phase
transition or a crossover. For example, Monte Carlo simulations have
found that the low frequency white noise in the energy and magnetization
is a maximum at the phase transition of the 2D ferromagnetic Ising model 
and the 2D 5-state Potts model \cite{zhichen}.
In addition, the maximum in the noise 
power of the defect density at low frequency occurs
at the onset of defect proliferation in a two-dimensional liquid 
modeled by a Yukawa potential~\cite{reichhardt2003, reichhardt2007}.

\begin{figure}
\begin{center}
\includegraphics[width=0.8 \textwidth]{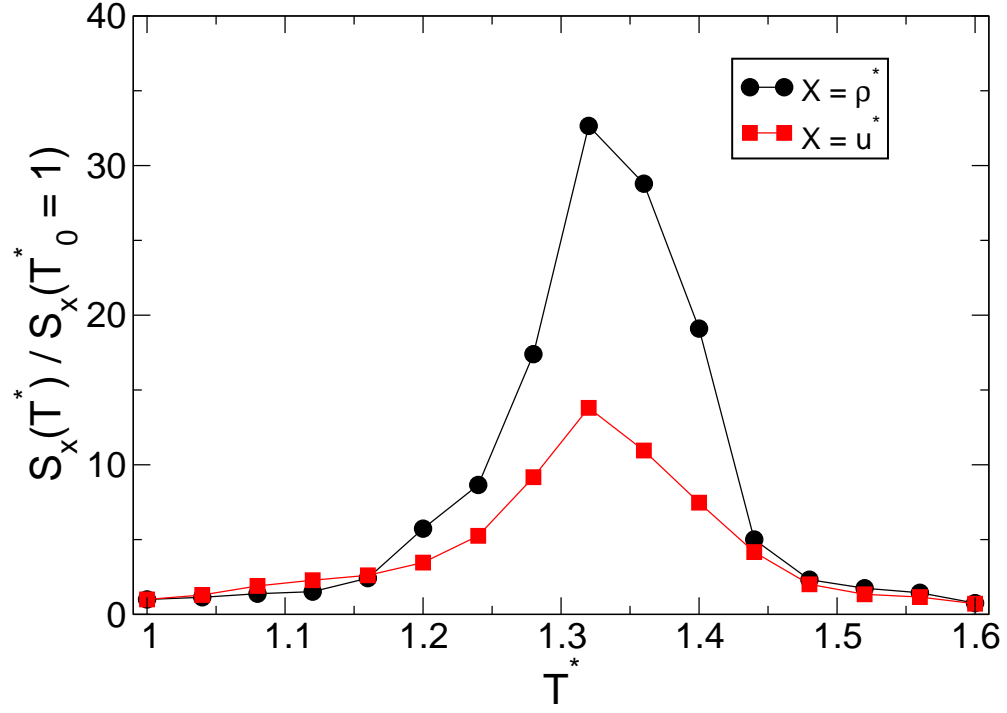}
\caption{\label{fig:spectralmag} (Color online) 
The magnitude of the normalized power spectra $S_{X}(T^{*}) / S_{X}(T^{*}_{0}=1)$ 
versus temperature $T^{*}$ for the density ($X = \rho^{*}$) and potential 
energy per particle ($X = u^{*}$) at low frequency at $P^{*}$ = 0.2. 
Both $S_{\rho^{*}}(T^{*})/S_{\rho^{*}}(T^{*}_{0} = 1)$ and 
$S_{u^{*}}(T^{*})/S_{u^{*}}(T^{*}_{0} = 1)$ exhibit maxima in the vicinity 
of the Widom line.} 
\end{center}
\end{figure}

We have found that the low frequency white noise in the density 
and the potential energy is a maximum at the Widom line.
In Figure \ref{fig:spectralmag}, we show values of the power spectra at 
low frequency.
We take the average value of the white noise part of the power spectrum 
$S_{X}$ where $X$ is $\rho^{*}$ or $u^{*}$ at each $T^{*}$. 
Each $S_{X}(T^{*})$ value in Fig.~\ref{fig:spectralmag} is normalized by 
$S_{X}(T^{*}_{0} = 1)$ to obtain the ratio $S_{X}(T^{*}) / S_{X}(T^{*}_{0} = 1)$.
For both the density ($X = \rho^{*}$) and potential energy ($X = u^{*}$) 
fluctuations, $S_{X}(T^{*}) / S_{X}(T^{*}_{0} = 1)$ has a maximum in the
vicinity of the Widom temperature $T^{*}_{\textrm{w}}$.
Therefore, our results show that a maximum in the low frequency
power can provide an additional signature of the Widom line.

\section{\label{sec:summary} Discussion and Summary}
We have performed molecular dynamics simulations of a Lennard-Jones
fluid in the supercritical region of the phase diagram to probe
the Widom line of the liquid-vapor phase transition and its effect on
the power spectra of the density and potential energy fluctuations. 

Extending from the liquid-vapor critical point, the Widom line, the locus of the 
maxima of the thermodynamic response functions, is a continuation of the 
liquid-vapor phase transition line.
The thermodynamic response functions in the supercritical region, 
such as the thermal expansion coefficient and the isobaric specific heat, 
have maxima along the Widom line.
Similar results have been found by studies of the Widom line of the 
hypothesized liquid-liquid phase transition in supercooled water
\cite{xu, kumar2006, kumar2007, kumar2008, stokely, abascal, wikfeldt}.

We studied the power spectra of the density and potential
energy fluctuations in the supercritical region.
Far away from the Widom line, the noise in the density fluctuations is white
at low frequencies and goes as $1/f^{2}$ at higher frequencies.
$1/f^2$ behavior is consistent with an
exponential relaxation process characterized by a single relaxation time.
In the vicinity of the Widom line we found that the density noise spectrum
could be divided into 3 frequency regimes:
$1/f^{0}$ at low frequency, $1/f^{1.2}$ at intermediate frequencies, and $1/f^{2}$
at high frequencies.
The intermediate region $1/f$ noise implies that there
is a distribution of relaxation times associated with the maxima in the response
functions along the Widom line. 
Eq.~(\ref{eqn:multiple}) implies that the narrowness of the 
frequency range where there is $1/f$ noise is due to a narrow distribution of
relaxation times. 
The narrow width of the intermediate region might be due to
the fact that density correlations are short lived at high temperatures.
In contrast to the density fluctuations, the power spectrum of the potential 
energy fluctuations along the Widom line does not exhibit three distinct 
frequency regimes.
Finally, we found that the low frequency white noise of the density and potential
energy fluctuations have their maxima along the Widom line. 
This suggests that noise power spectra, which have been used to probe
phase transitions \cite{zhichen}, can also be used to locate the Widom line. 

\begin{acknowledgements}
We thank Jaegil~Kim, Pradeep~Kumar, Albert~Libchaber and H.~Eugene~Stanley for helpful 
discussions. This work was supported by DOE grant DE-FG02-04ER46107.
\end{acknowledgements}

\newpage

\end{document}